\documentclass[twocolumn,noshowpacs,preprintnumbers,secnumarabic,amsmath,amssymb,nofootinbib]{revtex4}

\usepackage{graphicx}
\usepackage{dcolumn}
\usepackage{bm}

\def\beq{\begin{equation}}
\def\eeq{\end{equation}}
\def\be{\begin{equation}}
\def\ee{\end{equation}}
\def\bea{\begin{eqnarray}}
\def\eea{\end{eqnarray}}
\def\d{{\rm d}}

\DeclareRobustCommand{\SkipTocEntry}[4]{}

\begin{document}

\preprint{PUPT-2214}

\vspace{5cm}
\title{A Microscopic Limit on Gravitational Waves from D-brane
Inflation}

\author{Daniel Baumann}
\email{dbaumann@princeton.edu}

\author{Liam McAllister}%
 \email{lmcallis@princeton.edu}
\affiliation{%
Department of Physics, Princeton University,
Princeton, NJ 08544}

\date{\today}

\begin{abstract}
We derive a microscopic bound on the maximal field variation of
the inflaton during warped D-brane inflation. By a result of Lyth, this
implies an upper limit on the amount of gravitational waves
produced during inflation.  We show that a detection at the level
$r > 0.01$ would falsify slow roll D-brane inflation.
In DBI inflation, detectable tensors may be possible in special compactifications, provided that $r$ decreases rapidly during inflation.
We also show that for the special case of DBI
inflation with a quadratic potential, current observational
constraints imply strong upper bounds on the five-form flux.
\end{abstract}



\maketitle


\section{\label{sec:intro} Introduction}

In the foreseeable future it may be possible to detect primordial
gravitational waves \cite{CMB} produced during inflation
\cite{Inflation}.  This would be a spectacular opportunity to
reveal physics at energy scales that are unattainable in
terrestrial experiments.  In light of this possibility, it is
essential to understand the predictions made by various
inflationary models for gravitational wave production.  As we
shall review, a result of Lyth \cite{LythBound} connects
detectably large gravitational wave signals to motion of the
inflaton over Planckian distances in field space.  It is
interesting to know when suitably flat potentials over such large
distances are attainable in string compactifications, allowing a
potentially observable tensor signal in the associated string
inflation models.  In this paper we analyze this issue for the
case of warped D-brane inflation models \cite{Dvali,
KKLMMT,TyeReview}, and use compactification constraints to derive
a firm upper bound on the inflaton field range in Planck units.

For slow roll warped brane inflation, our result implies that the
gravitational wave signal is undetectably small.  This constraint
is model-independent and holds for any slow roll potential.  For
DBI inflation \cite{DBI,DBI2}, the limit on the field range forces the tensor signal to
be much smaller than the current observational bound. Detection in
a future experiment may be possible only if $r$ decreases rapidly
soon after scales observable in the cosmic microwave background
(CMB) exit the horizon.  This does occur in some models, but it
has a striking correlate: the scalar spectrum will typically have
a strong blue tilt and/or be highly non-Gaussian during the same epoch.

We also consider compactification constraints on the special case
of DBI inflation in which the potential is quadratic. We find that
observational constraints, together with our bound on the field
range, exclude scenarios with a large amount of five-form flux.
For a DBI model realized in a warped cone over an Einstein
manifold $X_5$, this translates into a very strong requirement on
the volume of $X_5$ at unit radius.  However, we show that
manifolds obeying this constraint do exist, at least in noncompact
models.  This translates the usual problem of accommodating a
large flux into the problem of arranging that $X_5$ has small
volume.

\section{\label{sec:lyth} The Lyth Bound}

In slow roll inflation the tensor fluctuation two-point function
is proportional to \beq { \frac{H^2}{M_P^2}}\, ,\eeq where $H$ is
the Hubble expansion rate.
 The scalar fluctuation two-point function is proportional to \beq \label{equ:tensor} H^2  \left( \frac{H}{\dot{\phi}} \right)^2\, .\eeq The first factor in (\ref{equ:tensor})
represents the two-point function of the scalar field, while the
second factor comes from the conversion of fluctuations of the
scalar field into fluctuations of the scale factor in the metric
(or scalar curvature fluctuations). This implies that the ratio
between the tensor and scalar two-point functions is proportional
to \beq \label{equ:TS}
 r \equiv 8\,  \Bigl( { \frac{\dot \phi}{H M_P}} \Bigr)^2 = 8\, \Bigl( {\frac{ d
\varphi}{d { \cal N} }} {\frac{1}{M_P}} \Bigr)^2\, , \eeq where
$\d {\cal N} = H \d t$ represents the differential of the number
of $e$-folds. We have fixed the numerical prefactor in
(\ref{equ:TS}) so that $r$ is defined as in the WMAP conventions.

This implies that the total field variation during inflation is
\beq \label{equ:lythINT} \frac{\Delta \varphi}{M_P} =
\frac{1}{8^{1/2}} \int_0^{{\cal{N}}_{\rm end}} \d {\cal{N}} \,
r^{1/2} \, , \eeq where ${\cal{N}}_{\rm end} \sim 60$ is the total
number of $e$-folds from the time the CMB quadrupole exits the
horizon to the end of inflation.

In any given model of inflation, $r$ is determined as a function
of ${\cal{N}}$. We therefore define \beq {\cal N}_{\rm eff} \equiv
\int_0^{{\cal N}_{\rm end}} \d {\cal N}\, \Bigl( \frac{r}{r_{\rm
CMB}} \Bigr)^{1/2}\, , \eeq so that \beq \frac{\Delta
\varphi}{M_P} = \Bigl( \frac{r_{\rm CMB}}{8} \Bigr)^{1/2} \, {\cal
N}_{\rm eff}\, . \eeq Here $r_{\rm CMB}$ denotes the
tensor-to-scalar ratio $r$ evaluated on CMB scales ($2 \le \ell
\lesssim 100$), $0 < {\cal N} < {\cal N}_{\rm CMB}  \approx 4$. We
use ${\cal N}_{\rm eff}$ to parameterize how far beyond ${\cal
N}_{\rm CMB}$ the support of the integral in (\ref{equ:lythINT})
extends.
If $r$ is precisely constant then ${\cal N}_{\rm eff} = {\cal
N}_{\rm end}$, if $r$ is monotonically increasing then ${\cal
N}_{\rm eff} > {\cal N}_{\rm end}$, and if $r$ decreases then
${\cal N}_{\rm eff} < {\cal N}_{\rm end}$.  For a detailed
discussion of related issues, see \cite{EKP}.

The {\it Lyth bound} \cite{LythBound} relates the maximal amount
of gravitational waves to the field variation $\Delta \varphi$
during inflation
 \beq \label{equ:lyth} r_{\rm CMB}  =
\frac{8}{({\cal{N}_{\rm eff}})^2} \Bigl( \frac{\Delta
\varphi}{M_P} \Bigr)^2\, . \eeq
The bound implies that a model producing a detectably large
quantity of gravitational waves necessarily involves field
variations of order the Planck mass. In the remainder of the paper
we will determine whether such large field variations are possible
in a class of string inflation models.

In slow roll inflation $r$ is proportional to the slow roll
parameter $\epsilon$. We can define a second slow roll parameter $\tilde \eta$
as the fractional variation of $\epsilon$ during one $e$-fold.
Then we have \beq \label{equ:dlnr} \frac{d \ln r}{d {\cal N}} =
\frac{d \ln \epsilon}{d {\cal N}} = \tilde \eta\, , \eeq where the
last equality is just the definition of $\tilde \eta$.
We can also write (\ref{equ:dlnr}) in terms of the spectral
indices of the scalar and tensor power spectra \bea \frac{d \ln
r}{d {\cal{N}}}
&=& n_T -(n_S-1) \nonumber \\
&=& -  \left[(n_S-1)+\frac{r}{8} \right]   \label{equ:rN}\, ,\eea
where we have used the usual single-field consistency condition
$n_T = - r/8$.

To determine ${\cal N}_{\rm eff}$, we notice that ${\cal N}_{\rm
eff} < {\cal N}_{\rm end}$ only if $\tilde \eta$ is negative.
Present observations \cite{WMAP3,SDSS} indicate that $|\tilde
\eta_{\rm CMB}|$ is very small on scales probed by CMB (${\cal N}
\lesssim 4$) and large-scale structure observations (${\cal N}
\lesssim 10$).  In particular, $\tilde \eta_{\rm CMB} \gtrsim
-0.03$.  Since the variation of $\tilde \eta$ is second order in
slow roll we may assume that $\tilde \eta$ remains small
throughout inflation. Integrating (\ref{equ:dlnr}), we find a
range ${\cal N}_{\rm eff} \sim 30 - 60$ in (\ref{equ:lyth}).
Nearly all of the range for $\tilde \eta_{\rm CMB}$ allowed by
WMAP3+SDSS \cite{WMAP3,SDSS} actually corresponds to ${\cal
N}_{\rm eff} \gtrsim 50$.  To get a conservative bound, we have
considered the most negative allowed values of $\tilde \eta_{\rm
CMB}$, corresponding to the largest allowed values of $n_{S}-1$
and $r$, and this gives ${\cal N}_{\rm eff} \sim 30$.  Direct
observation of gravitational waves by some futuristic
gravitational wave detector such as the Big Bang Observer (BBO)
would put a similar lower bound on ${\cal N}_{\rm eff}$ (see {\it
e.g.} \cite{LathamBBO}).

With this input, the Lyth bound for slow roll models of inflation becomes
 \beq \label{equ:lythEXT} r_{\rm CMB} \lesssim
\frac{8}{30^2} \Bigl( \frac{\Delta \varphi}{M_P} \Bigr)^2\, . \eeq

\section{Constraint on Field Variation in Compact Spaces}

In this section we determine the maximum field range of the
inflaton in warped D-brane inflation.\footnote{The implications of
field range limits for eternal D-brane inflation have been
discussed in \cite{ChenEternal}.}  By (\ref{equ:lyth}) or
(\ref{equ:lythEXT}), this will imply a model-independent upper
limit on gravitational wave production in this scenario.

\addtocontents{toc}{\SkipTocEntry}
\subsection{Warped Throat Compactifications}

Consider a warped flux compactification of type IIB string theory
to four dimensions \cite{FluxReview}, with the line element \beq
\label{equ:lineelement} \d s^2 = h^{-1/2}(y) g_{\mu\nu} \d x^{\mu} \d
x^{\nu} + h^{1/2}(y)g_{ij} \d y^{i} \d y^{j} \, . \eeq  We will be
interested in the case that the internal space has a conical
throat, {\it{i.e.}} a region in which the metric is locally of the
form\footnote{We use $\rho$ to denote the radial direction,
because the conventional symbol $r$ is already in use.} \beq
g_{ij}\d y^{i} \d y^{j} = \d \rho^2 + \rho^2 \d s_{X_5}^2\, ,\eeq
for some five-manifold $X_5$. The metric on this cone is
Calabi-Yau provided that $X_5$ is a Sasaki-Einstein space. If the
background contains suitable fluxes, the metric in the throat
region can be highly warped.

Many such warped throats can be approximated locally, {\it{i.e.}}
for a small range of $\rho$, by the geometry $AdS_5 \times X_5$,
with the warp factor
\beq \label{equ:hAdS} h(\rho) = \Bigl( \frac{R}{\rho} \Bigr)^4\, ,
\eeq where $R$ is the radius of curvature of the $AdS$ space.  In
the case that the background flux is generated entirely by $N$
dissolved D3-branes placed at the tip of the cone, we have the
relation \cite{Gubser} \beq \frac{R^4}{(\alpha')^2} = 4 \pi g_s N
\frac{\pi^3}{{\rm Vol}(X_5)}\, . \eeq Here ${\rm Vol}(X_5)$
denotes the dimensionless volume of the space $X_5$ with unit
radius.\footnote{An equivalent definition of ${\rm Vol}(X_5)$,
which may be more clear when it is difficult to define a radius,
is as the angular factor in the integral defining the volume of a
cone over $X_5$.} Generically, we expect this volume to obey ${\rm
Vol}(X_5) = {\cal O}(\pi^3)$, {\it e.g.} ${\rm Vol}(S^5) = \pi^3$,
${\rm Vol}(T^{1,1}) = \frac{16}{27} \pi^3$. However, very small
volumes are possible, for example by performing orbifolds.

Warped throats have complicated behavior both in the infrared and
the ultraviolet.  For almost all $X_5$, no smooth tip geometry,
analogous to that of the Klebanov-Strassler throat \cite{KS}, is
known. Furthermore, the ultraviolet end of the throat, where the
conical metric is supposed to be glued into a compact bulk, is
poorly understood.  These regions are geometric realizations of
what are called the `IR brane' and `Planck brane' in
Randall-Sundrum models.
In this note we study constraints that are largely independent of
the properties of these boundaries.  We take the throat to extend
from the tip at $\rho=0$ up to a radial coordinate $\rho_{UV}$,
where the ultraviolet end of the throat is glued into the bulk of
the compactification. Background data, in particular three-form
fluxes, determine $\rho_{UV}$, but we will find that $\rho_{UV}$
cancels from the quantities of interest.

To summarize our assumptions: we consider a throat that is
a warped cone over some
Einstein space $X_5$, but may have complicated modifications in
the infrared and ultraviolet.  This very large class of geometries
includes the backgrounds most often studied for warped brane
inflation, but it would be interesting to understand even more
general warped throats.
\addtocontents{toc}{\SkipTocEntry}
\subsection{A Lower Bound on the \\Compactification Volume}

Standard dimensional reduction gives the following relation
between the four-dimensional Planck mass $M_P$, the warped volume
of the compact space $V_6^w$, the inverse string tension
$\alpha'$, and the string coupling $g_s$: \beq M_P^2 \equiv \frac{
V_6^w}{\kappa_{10}^2}\, , \eeq where $\kappa_{10}^2 \equiv
\frac{1}{2} (2 \pi)^7 g_s^2 (\alpha')^4$. The warped volume of the
internal space is \beq V_6^w = \int \d^6 y \sqrt{ g}\,  h\, . \eeq
Formally this may be split into separate contributions from the
bulk and the throat region \beq V_6^w \equiv ( V_6^w)_{\rm bulk} +
( V_6^w)_{\rm throat}\, . \eeq The throat contribution is \bea
(V_6^w)_{\rm throat} &\equiv& {\rm Vol}(X_5) \int_0^{\rho_{UV}} \d \rho \, \rho^5 h(\rho) \nonumber \\
&=& \frac{1}{2}   {\rm Vol}(X_5) R^4 \rho_{UV}^2 \nonumber \\
&=& 2\pi^4 g_{s} N (\alpha')^2\, \rho_{UV}^2 \, .
\label{equ:Vthroat} \label{equ:throat} \eea A key point is that
the {\it{warped}} throat volume is independent of ${\rm
Vol}(X_5)$.

The result (\ref{equ:throat}) is rather robust.  To confirm that
(\ref{equ:hAdS}) is a suitable approximation for the warp factor,
we note that in the Klebanov-Tseytlin regime \cite{KT} of a
Klebanov-Strassler throat, the warp factor may be written as \beq
\label{equ:KT} h(\rho) = \Bigl( \frac{L}{\rho} \Bigr)^4 \ln
\frac{\rho}{\rho_s}\, , \eeq where $L^4 \equiv \frac{3}{2 \pi}
\frac{g_s M^2}{N} R^4$, $\ln \frac{\rho_{UV}}{\rho_s} \approx
\frac{2 \pi }{3} \frac{K}{g_s M}$ and $N \equiv M K$. Integrating
(\ref{equ:KT}) one finds \beq
 (V_6^w)_{\rm throat} = \frac{1}{2} {\rm Vol}(T^{1,1}) R^4 \rho_{UV}^2\, ,
 \eeq
in agreement with equation (\ref{equ:Vthroat}).

The bulk volume is model-dependent, but we can impose a very
conservative lower bound on the total warped volume by omitting
the bulk volume, \beq V_6^w
> (V_6^w)_{\rm throat}\, . \eeq This implies a lower limit on the
four-dimensional Planck mass in string units \beq M_P^2 >
\frac{(V_6^w)_{\rm throat}}{\kappa_{10}^2}\, . \eeq

\addtocontents{toc}{\SkipTocEntry}
\subsection{An Upper Bound on the Field Range}

Let us now consider inflation driven by the motion of a D3-brane
in the background (\ref{equ:lineelement}).  The
canonically-normalized inflaton field is \beq \varphi^2 = T_3
\rho^2 \, , \quad T_3 \equiv \frac{1}{(2\pi)^3} \frac{1}{g_s
(\alpha')^2}\, . \eeq

The maximal radial
displacement of the brane in the throat is the length of the
throat, from the tip $\rho_{IR} \approx 0$ to the ultraviolet end,
$\rho_{UV}$,
so  that $\Delta \rho \lesssim \rho_{UV}$.
Naively one could think that the range of the inflaton could be
made arbitrarily large by increasing the length of the throat.
However, what is relevant is the field range in four-dimensional
Planck units, which is
\beq \Bigl( \frac{\Delta \varphi}{M_P} \Bigr)^2 < \frac{T_3
\rho_{UV}^2}{M_P^2} < \frac{T_3 \kappa_{10}^2 \rho_{UV}^2}{(V_6^w)_{\rm
throat}}\, . \eeq Substituting equation (\ref{equ:throat}) gives
the following important constraint on the maximal field variation
in four-dimensional Planck units \beq \label{equ:result}
 \Bigl( \frac{\Delta \varphi}{M_P} \Bigr)^2 <
\frac{4}{N} \, . \eeq

Two comments on this result are in order. First, the field range
in Planck units only depends on the background charge $N$ and is manifestly independent of the choice of $X_5$,
so our result is the same for any throat that is a warped cone over
some $X_5$.  Second, the size of the throat, and hence the
validity of a supergravity description of the throat, increases
with $N$. In the same limit, the field range in Planck units
decreases, because the large throat volume causes the
four-dimensional Planck mass to be large in string units. Because
$N=0$ corresponds to an unwarped throat, we require at the very
least $N \ge 1$; in practice, $N \gg 1$ is required for a
controllable supergravity description.

The bound (\ref{equ:result}) is extremely conservative, because we
have neglected the bulk volume, which in many cases will actually
be larger than the throat volume.  Modifications of the geometry
at the tip of the throat, where $\rho \ll R$, provide negligible
additional field range.  One might also try to evade this bound by
considering a stack of $n$ D3-branes moving down the throat, which
increases the effective tension.  However, the backreaction from
such a stack is important unless $n \ll N$, so this will not
produce a bound weaker than (\ref{equ:result}) with $N=1$.

\section{Implications for Slow Roll Brane Inflation}

Via the Lyth relation (\ref{equ:lyth}), the bound
(\ref{equ:result}) translates into a microscopic constraint on the
maximal amount of gravitational waves produced during warped brane
inflation \beq \label{equ:generalslowrollbound} \frac{r_{\rm
CMB}}{0.009} \le
 \frac{1}{N} \Bigl( \frac{60}{{\cal N}_{\rm eff}}\Bigr)^2\, . \eeq As
explained in \S\ref{sec:lyth}, for slow roll inflation,
recent observations \cite{WMAP3, SDSS} imply \beq
\label{equ:neffrange} 30 \lesssim {\cal{N}_{\rm eff}} \lesssim
{\cal N}_{\rm end} \approx 60\, . \eeq Let us stress that the lower
part of this range is occupied only by models with a large,
positive scalar running or a blue scalar spectrum and a large
tensor fraction.

This implies \beq \label{equ:extendedslowrollbound} \frac{r_{\rm
CMB}}{0.009} \lesssim \frac{4}{N} \, . \eeq Near-future CMB
polarization experiments \cite{CMB, Clover} will
probe\footnote{The ultimate detection limit is probably around $r
\sim 10^{-3}-10^{-4}$. Measuring even lower $r$ is prohibited by
the expected magnitude of polarized dust foregrounds and by the
lensing conversion of primordial E-modes to
B-modes.\cite{SpergelPrivate}} $r_{\rm CMB} \gtrsim {\cal
O}(10^{-2})$. Detection of gravitational waves in such an
experiment would therefore imply that $N<4$.
This implies that the space is effectively unwarped, and that the
supergravity description is uncontrolled. We therefore find that
warped D-brane inflation can be falsifed by a detection of
gravitational waves at the level $r_{\rm CMB} \gtrsim 0.01$.

One might have anticipated this result on the grounds that D-brane
inflation models are usually considered of the `small-field' type,
and are typically thought to predict an unobservably small tensor
fraction. Let us stress, however, that extracting precise
predictions from D-brane inflation scenarios is rather involved,
and requires careful consideration, and fine-tuning, of the
potential introduced by moduli stabilization \cite{BDKMMM}. It is
quite unlikely that the fully corrected potential will enjoy the
same exceptional flatness as the uncorrected potential given in
\cite{KKLMMT}.  As moduli stabilization effects increase
$\epsilon$, they increase $r$, and {\it{a priori}} this may be
expected to lead to observable gravitational waves. Indeed, it has
been argued in the context of more general single-field inflation
that minimally tuned models correlate with maximal gravitational
wave signals \cite{Latham}. Nevertheless, our result implies that
even the {\it{maximal}} signal in warped D-brane inflation is
undetectably small. We have thus excluded the possibility of
detectable tensors on purely kinematic grounds, {\it{i.e.}} by
using only the size of the field space.

\section{Implications for DBI Inflation}

A very interesting alternative to slow roll inflation arises when
nontrivial kinetic terms drive inflationary expansion.
The DBI
model \cite{DBI,DBI2,ChenIR,otherDBI} is a string theory
realization of this possibility in which a D3-brane moves rapidly
in a warped background.  In this section we combine the Lyth bound
with our field range bound (\ref{equ:result}) to constrain the
tensor signal in DBI inflation.

\addtocontents{toc}{\SkipTocEntry}
\subsection{A Generalized Lyth Bound}

We present the Lyth bound in a theory with a general kinetic term, then specialize to the DBI case.
Consider the action \cite{Garriga:1999vw} \beq \label{equ:actionP}
S = \frac{1}{2} \int \d^4 x \sqrt{-g} \Bigl[ M_P^2 {\cal R} + 2
P(X, \varphi)\Bigr]\, , \eeq
 where $P(X,\varphi)$ is a general
function of the inflaton $\varphi$ and of $X \equiv - \frac{1}{2}
g^{\mu \nu} \partial_\mu \varphi \partial_\nu \varphi$. For slow
roll inflation \beq P(X, \varphi) \equiv X - V(\varphi)\, , \eeq
while DBI inflation may be parameterized by \cite{DBI} \beq
\label{equ:dbiaction} P(X,\varphi) \equiv - f^{-1}(\varphi)
\sqrt{1- 2 f(\varphi) X} + f^{-1}(\varphi) - V(\varphi)\, , \eeq
 where
$f^{-1}(\varphi) = T_3 h^{-1}(\varphi)$ is the rescaled warp factor. From (\ref{equ:actionP})
we find the energy density in the field to be $\rho = 2 X
P_{,X}-P$. We also define the speed of sound as \beq c_s^2 \equiv
\frac{dP}{d\rho} = \frac{P_{,X}}{\rho_{,X}} =
\frac{P_{,X}}{P_{,X}+2X P_{,XX}}\, . \eeq We define slow variation
parameters in analogy with the standard slow roll parameters \bea
\label{equ:eps}
\epsilon &\equiv& - \frac{\dot H}{H^2} = \frac{X P_{,X}}{M_P^2 H^2}\, ,\\
\tilde \eta &\equiv & \frac{\dot \epsilon}{\epsilon H}\, ,\\
s &\equiv& \frac{\dot{c_s}}{c_s H}\, . \eea To first order in
these parameters the basic cosmological observables are \bea
P_S &=& \frac{1}{8 \pi^2 M_P^2} \frac{H^2}{c_s \epsilon}\, ,\\
P_T &=& \frac{2}{\pi^2} \frac{H^2}{M_P^2}\, ,\\
n_S -1 &=& - 2 \epsilon - \tilde \eta - s\, ,\\
n_T &=& -2 \epsilon\, . \eea The tensor-to-scalar ratio in these
generalized inflation models is \beq \label{equ:r} r = 16\,  c_s
\epsilon\, . \eeq  This nontrivial dependence on the speed of
sound implies a modified consistency relation \beq \label{equ:mce}
r = - 8 \, c_s n_T\, . \eeq As discussed recently by Lidsey and
Seery \cite{Lidsey:2006ia} for the case of DBI inflation, equation
(\ref{equ:mce}) provides an interesting possibility of testing
fast roll inflation . The standard slow roll predictions
are recovered in the limit $c_s = 1$.

Restricting to the homogeneous mode $\varphi(t)$ we find from
(\ref{equ:eps}) that \beq \frac{\d \varphi}{M_P} = \sqrt{\frac{2
\epsilon}{P_{,X}}} \, \d {\cal{N}} \eeq and hence \beq
\frac{\Delta \varphi}{M_P}= \int_0^{{\cal{N}}_{\rm end}}
\sqrt{\frac{r}{8} \frac{1}{c_s P_{,X}}} \, \d {\cal{N}}\, . \eeq
Notice the nontrivial generalization of the slow roll result
(\ref{equ:lythINT}) through the factor $c_s P_{,X}$. For DBI
inflation this factor happens to be \beq c_s P_{,X} = 1\, , \eeq
where \beq c_s^2 = 1- 2 f(\varphi) X \equiv
\frac{1}{\gamma^2(\varphi)}\, , \eeq so that the Lyth bound
remains the same as for slow roll inflation.\footnote{This result
has been obtained independently by S. Kachru and by H. Tye.}

The variation of $r$ during inflation follows from (\ref{equ:r})
\beq \label{equ:rNDBI1} \frac{d \ln r }{d {\cal{N}}} =  \frac{d \ln \epsilon}{d
{\cal{N}}} + \frac{d \ln c_s}{d {\cal{N}}}  =  \tilde \eta + s\, .
\eeq
While the observed near scale-invariance of the density
perturbations restricts the magnitude of  $s = d \ln c_s/ d {\cal
N}$ in the range $0 \le {\cal N} \lesssim 10$, outside that window
$s$ can in principle become large and negative. By
(\ref{equ:rNDBI1}) this would source a rapid decrease in $r$.
Note, however, that the results given in this section are first
order in $s$ and so receive important corrections when $s$ is
large. Furthermore, omission of terms in the DBI action involving two
or more derivatives of $\varphi$ may not be consistent when $s$ is
sufficiently large.

Constraints on the evolution of $r$ may also be understood by rewriting equation (\ref{equ:rNDBI1}) as
\bea \frac{d \ln r}{d {\cal{N}}}
&=& n_T -(n_S-1)\nonumber \\
&=& -  \left[(n_S-1)+\frac{r}{8 c_s} \right]
 \label{equ:rNDBI}
\, .\eea
This implies that $r$ can decrease significantly only if the scalar spectrum becomes very blue ($n_S - 1 > 0$) and/or the speed of sound becomes very small, so that $r/c_s$ is large.  During the time when observable scales exit the horizon this possibility is significantly constrained, but outside that window $r$ may decrease rapidly in some models.

\addtocontents{toc}{\SkipTocEntry}
\subsection{Constraints on Tensors}

Just as in slow roll inflation we can write
 \beq \label{equ:DBIBound} \frac{r_{\rm CMB}}{0.009} \le
 \frac{1}{N} \Bigl( \frac{60}{{\cal N}_{\rm eff}}\Bigr)^2\, . \eeq
However, in DBI inflation we have to allow for the possibility
that a nontrivial evolution of the speed of sound allows ${\cal
N}_{\rm eff}$ to be considerably smaller than ${\cal N}_{\rm
end}$, which weakens the Lyth
bound. The precise value of ${\cal N}_{\rm eff}$ will be highly model-dependent.\\

In light of the constraint (\ref{equ:DBIBound}), constructing a successful DBI model
with detectable tensors is highly nontrivial.  First of all, such
a model must produce a spectrum of scalar perturbations consistent with
observations, {\it{i.e.}} with the appropriate amplitude and with
a suitably small level of non-Gaussianity. Then, the model should include
each of the following {\it{additional}} elements related to the
large tensor signal:

\begin{enumerate}
\item A consistent compactification in which
\beq (V_6^w)_{\rm bulk} \ll (V_6^w)_{\rm throat} \, , \eeq so that
the inequality in (\ref{equ:result}) may be nearly saturated.
\item A {\it{small}} five-form flux $N$, together with a demonstration that the supergravity corrections and brane backreaction are under
control in this difficult limit.
\item A decrease in $r$ that is rapid enough to ensure that ${\cal{N}_{\rm eff}} \ll 30$.
In this situation the slow variation parameters $\tilde{\eta}$,
$s$ cannot both be small, substantially complicating the analysis.

\end{enumerate}

It would be extremely interesting to find a system that satisfies
all these constraints, especially because this would be a rare
example of a complete string inflation model with detectable
tensors.

\addtocontents{toc}{\SkipTocEntry}
\subsection{Constraints on Quadratic DBI Inflation}

In this section we illustrate our considerations for one important
class of DBI models, those with a quadratic potential.\footnote{We
consider the so-called `UV model', {\it{i.e.}} with a D3-brane
moving toward the tip of the throat; {\it{cf.}} \cite{ChenIR} for
an interesting alternative.}

We consider an action of the form (\ref{equ:dbiaction}), with \beq
V(\varphi) = \frac{1}{2}m^2\varphi^2 \, .\eeq At sufficiently late
times, the Hubble parameter is \cite{DBI}
\beq H(\varphi) = c \, \varphi \, , \eeq for some constant $c$.
Using this in (\ref{equ:eps}), one finds \beq \label{equ:UVgamma}
\epsilon\, \gamma(\varphi) = 2 M_P^2 \Bigl(
\frac{H_\varphi}{H}\Bigr)^2 = 2
\left(\frac{M_P}{\varphi}\right)^2\, . \eeq This relates the DBI
Lorentz factor $\gamma$ to the slow roll parameter $\epsilon < 1$
and to the inflaton field value.

\addtocontents{toc}{\SkipTocEntry}
\subsubsection{Microscopic Constraint from Limits on Non-Gaussianity}

Observational tests of the non-Gaussianity of the primordial
density perturbations are most sensitive to the three-point
function of the comoving curvature perturbations. It is usually
assumed that the three-point function has a form that would follow
from the field redefinition \beq \zeta = \zeta_g - \frac{3}{5}
f_{NL} \zeta_g^2\, , \eeq where $\zeta_g$ is Gaussian. The scalar
parameter $f_{NL}$ quantifies the amount of non-Gaussianity.
 It is a function of three momenta which form a triangle in Fourier space.
 Here we cite results for the limit of an equilateral triangle. Slow
roll models predict $f_{NL} \ll 1$ \cite{malda}, which is far
below the detection limit of present and future observations. For
generalized inflation models represented by the action
(\ref{equ:actionP}) one finds \cite{Chen:2006nt} \beq
\label{equ:fNLP} f_{NL} = \frac{35}{108} \Bigl( \frac{1}{c_s^2} -1
\Bigr) - \frac{5}{81} \Bigl( \frac{1}{c_s^2} -1 - 2
\Lambda\Bigr)\, , \eeq where \beq \Lambda \equiv \frac{X^2 P_{,XX}
+ \frac{2}{3} X^3 P_{,XXX}}{X P_{,X} + 2 X^2 P_{,XX}}\, . \eeq

For the case of DBI inflation (\ref{equ:dbiaction}), the second
term in (\ref{equ:fNLP}) is identically zero \cite{Chen:2006nt},
so that the prediction for the level of non-Gaussianity
(\ref{equ:fNLP}) is\footnote{Notice that this result is generic to
DBI inflation and is independent of the choice of the potential and
the warp factor. This is in contrast to other observables like
$n_S$, $P_S$, {\it{etc}}.}
\beq f_{NL} = \frac{35}{108} \Bigl( \frac{1}{c_s^2} - 1\Bigr)
\approx \frac{35}{108}\gamma^2\, , \eeq where the second relation
holds when $c_{s} \ll 1$.  This result leads to an upper bound on
$\gamma$ from the observed limit on the non-Gaussianity of the
primordial perturbations.  The recent analysis of the WMAP3
\cite{WMAP3} data in \cite{HarvardfNL} gives $ -256 < f_{NL} < 332
$ (95\% confidence level), which implies
\beq \label{equ:gammabound} \gamma \lesssim 32 \, . \eeq
Using the expressions (\ref{equ:r}) and (\ref{equ:UVgamma}), we
have \beq N < 4 \Bigl( \frac{M_P}{\varphi} \Bigr)^2 = \frac{r
\gamma^2}{8} = \frac{27}{70} r f_{NL} \, .\eeq Combining the
observational bound on gravitational waves \cite{WMAP3}, $r <
0.3$, with the bound on non-Gaussianity, we find \beq
\label{equ:x1} N \lesssim 38\, . \eeq Quadratic DBI inflation with
a larger amount of five-form flux is hence excluded by current
observations. The Planck satellite may be sensitive enough to give
the limits $|f_{NL}| < 50$ \cite{PlanckfNL} and $r < 0.05$.
Non-observation at these levels would give the bound $N < 1$,
excluding quadratic DBI inflation.

\addtocontents{toc}{\SkipTocEntry}
\subsubsection{Microscopic Constraint from the Amplitude of Primordial Perturbations}

Garriga and Mukhanov \cite{Garriga:1999vw} have derived the
perturbation spectrum for theories with non-canonical kinetic
terms \beq \left. P_S = \frac{1}{8 \pi^2 M_P^2} \frac{H^2}{c_s
\epsilon}\right|_{c_s k = aH}\, . \eeq

Using $c_s^{-1} = \gamma(\varphi) = \sqrt{1 + 4 M_P^4 f(\varphi) (H_\varphi)^2}$ for DBI inflation, this becomes \cite{DBI, DBI2}
\beq P_S = \frac{16}{\pi^2} \frac{\gamma^2 (\gamma^2-1)}{(r
\gamma^2)^2} \frac{1}{M_P^4 f}\, . \eeq For the $AdS_5$ warp
factor (\ref{equ:hAdS}) \beq M_P^4 f(\varphi) = \lambda \Bigl(
\frac{M_P}{\varphi} \Bigr)^4 = \lambda \Bigl( \frac{r
\gamma^2}{32} \Bigr)^2\, , \eeq where \beq
 \lambda \equiv T_3 R^4 = \frac{\pi}{2} \frac{N}{{\rm Vol}(X_5)}\, ,
\eeq we find
\beq \label{equ:62} P_S = \Bigl( \frac{32}{\pi}  \Bigr)^3 \times
\frac{\gamma^2 (\gamma^2-1)}{(r \gamma^2)^4} \frac{{\rm
Vol}(X_5)}{N}\, . \eeq In the relativistic limit we have $\gamma^2
(\gamma^2 -1) \sim \gamma^4 = 9 f_{NL}^2$ and (\ref{equ:62})
becomes \beq P_S =  \Bigl( \frac{32}{3 \pi}  \Bigr)^3 \frac{3}{r^4
f_{NL}^2} \, \frac{{\rm Vol}(X_5)}{N} \gtrsim  0.1\, \frac{{\rm
Vol}(X_5)}{N}\, , \eeq
where the last relation comes from the current observational bounds on $r$
and $f_{NL}$ on CMB scales.
COBE or WMAP give the normalization $P_S \approx 10^{-9}$, so that
we arrive at the condition \beq \label{equ:x2} N \gtrsim 10^{8} \,
{\rm Vol}(X_5) \, . \eeq

The requirements (\ref{equ:x1}) and (\ref{equ:x2}) are clearly
inconsistent for the generic case, ${\rm Vol}(X_5) \sim {\cal
O}(\pi^3)$.  We conclude that quadratic DBI inflation in warped
throats\footnote{Throats that are not cones over Einstein
manifolds could evade this constraint, and may be a more natural
setting for realizing the DBI mechanism.  We thank E. Silverstein
for explaining this to us.} cannot simultaneously satisfy the
observational constraints on the amplitude and Gaussianity of the
primordial perturbations unless ${\rm Vol}(X_5) \lesssim 10^{-7}$.  In
particular, this excludes realization of this scenario in a
cut-off $AdS$ model or in a Klebanov-Strassler \cite{KS} throat.
Cones with very small values of ${\rm Vol}(X_5)$ can be
constructed by taking orbifolds or considering $Y^{p,q}$ spaces in
the limit that $q $ is fixed and $p\to \infty$
\cite{Gauntlett:2004yd}. However, it seems rather unlikely that
one could embed these spaces into a string compactification.

\section{Conclusions}

We have established a firm upper bound on the canonical field
range in Planck units for a D3-brane in a warped throat.  This
range can never be large, and can be of order one only in the
limit of an unwarped throat attached to a bulk of negligible
volume.  Combined with the Lyth relationship \cite{LythBound}
between the variation of the inflaton field during inflation and
the gravitational wave signal, this implies a constraint on the
tensor fraction in warped D-brane inflation.  The tensor signal is
undetectably small in slow roll warped D-brane inflation,
regardless of the form of the potential. In DBI inflation,
detectable tensors may be possible only in a poorly-controlled limit of
small warping, moderately low velocity, rapidly-changing speed of
sound, and substantial backreaction.
In this case, the scalar
spectrum will typically have a strong blue tilt and/or become highly non-Gaussian shortly after
observable scales exit the horizon.

We have also presented stronger constraints for the case of DBI
inflation with a quadratic potential, finding that combined
observational constraints on tensors and non-Gaussianity imply an
upper bound $N \lesssim 38$ on the amount of five-form flux.
Near-future improvements in the experimental limits could imply $N
< 1 $ and thus exclude the model.  For models realized in a warped
cone over a five-manifold $X_5$, current limits imply that the
dimensionless volume of $X_5$, at unit radius, is smaller than
$10^{-7}$.  Manifolds of this sort do exist; extremely high-rank
orbifolds and cones over special $Y^{p,q}$ manifolds are examples,
but it is not clear that these can be embedded in a string
compactification.

Although our result resonates with some well-known effective field
theory objections (see {\it{e.g.}} \cite{LythRiotto}) to
controllably flat inflaton potentials involving large field
ranges, we stress that our analysis was entirely explicit and did
not rely on notions of naturalness or of fine-tuning.

Our microscopic limit on the evolution of the inflaton implies
that a detection of primordial gravitational waves would rule out
most models of warped D-brane inflation, and place severe pressure
on the remainder. We expect that compactification constraints on
canonical field ranges imply similar bounds in many other string
inflation models \cite{Kahler}.\footnote{We should mention one
promising string inflation scenario, N-flation \cite{nflation},
that does predict observable tensors.  It would be very
interesting to understand whether this model can indeed be
realized in a string compactification.}   In this sense, current
models of string inflation do not readily provide detectable
gravitational waves \cite{ShamitTalk}. However, this is not yet by
any means a firm prediction of string theory, and it is more
important than ever to search for a compelling model of
large-field string inflation that overcomes this obstacle.  Given the
apparent difficulty of achieving super-Planckian field variations
with controllably flat potentials for scalar fields in string
theory, a detection of primordial gravitational waves would
provide a powerful selection principle for string inflation models
and give significant clues about the fundamental physics
underlying inflation.


\addtocontents{toc}{\SkipTocEntry}
\section*{Acknowledgments}
We thank Igor Klebanov and Juan Maldacena for initial
collaboration and many insightful comments.  We are grateful to
Richard Easther, Shamit Kachru, Eva Silverstein, and Paul
Steinhardt for comments on a draft, and we thank Latham Boyle,
Cliff Burgess, Chris Herzog, Nissan Itzhaki, Hiranya Peiris,
Fernando Quevedo, Gary Shiu, Andrew Tolley, Henry Tye, and Herman
Verlinde for helpful discussions. We thank Cliff Burgess and the
Perimeter Institute for their kind hospitality while this work was
completed. L.M. also thanks the organizers of the KITP workshop on
String Phenomenology. This research was supported in part by the
United States Department of Energy, under contract
DE-FG02-90ER-40542.


\begingroup\raggedright\endgroup

\end{document}